\documentclass[11pt,a4paper]{emulateapj}

\usepackage{epsfig}
\usepackage{amsmath}

\def\apj{ApJ}
\def\apjl{ApJL}
\def\mnras{MNRAS}

\def\apjs{ApJS}

\def\gs{\mathrel{\raise0.35ex\hbox{$\scriptstyle >$}\kern-0.6em\lower0.40ex\hbox{{$\scriptstyle \sim$}}}} 
\def\ls{\mathrel{\raise0.35ex\hbox{$\scriptstyle <$}\kern-0.6em\lower0.40ex\hbox{{$\scriptstyle \sim$}}}}

\def\Wm2{\,\hbox{W}\,\hbox{m}^{-2}} 
\def\gsim{\mathrel{\raise0.35ex\hbox{$\scriptstyle >$}\kern-0.6em\lower0.40ex\hbox{{$\scriptstyle \sim$}}}} 
\def\lsim{\mathrel{\raise0.35ex\hbox{$\scriptstyle <$}\kern-0.6em\lower0.40ex\hbox{{$\scriptstyle \sim$}}}} 
\def\ltsima{$\; \buildrel < \over \sim \;$} 
\def\simlt{\lower.5ex\hbox{\ltsima}} 
\def\gtsima{$\; \buildrel > \over \sim \;$} 
\def\simgt{\lower.5ex\hbox{\gtsima}}

\lefthead{Swinbank et al.}  \righthead{Star Forming Regions in SDP\,81 at $z$\,=\,3.04 from ALMA}

\begin{document}

\title {ALMA maps the Star-Forming Regions in a Dense Gas Disk at
  \lowercase{$z$}\,$\sim$\,3}
  
\author{
A.\,M.\ Swinbank,\altaffilmark{1,2}
S.\ Dye,\altaffilmark{3}
J. W. Nightingale,\altaffilmark{3}
C.\ Furlanetto,\altaffilmark{3,4}
Ian Smail,\altaffilmark{1,2}
A.\ Cooray,\altaffilmark{5}
H.\ Dannerbauer,\altaffilmark{6}
L.\ Dunne,\altaffilmark{7,8}
S.\ Eales,\altaffilmark{9}
R.\ Gavazzi,\altaffilmark{10}
T.\ Hunter,\altaffilmark{11}
R.\, J.\ Ivison,\altaffilmark{8,12}
M.\ Negrello,\altaffilmark{13}
I.\ Oteo,\altaffilmark{8,12}
R.\ Smit,\altaffilmark{1,2}
P.\ van der Werf,\altaffilmark{14}
C.\ Vlahakis,\altaffilmark{15,16}}
\setcounter{footnote}{0}
\altaffiltext{1}{Institute for Computational Cosmology, Department of Physics, Durham University, South Road, Durham DH1 3LE, UK; email: a.m.swinbank@dur.ac.uk}
\altaffiltext{2}{Center for Extragalactic Astronomy, Department of Physics, Durham University, South Road, Durham DH1 3LE, UK}
\altaffiltext{3}{School of Physics and Astronomy, Nottingham University, University Park, Nottingham, NG7 2RD, UK}
\altaffiltext{4}{CAPES Foundation, Ministry of Education of Brazil, Bras\'ilia/DF, 70040-020, Brazil}
\altaffiltext{5}{Astronomy Department, California Institute of Technology, MC 249-17, 1200 East California Boulevard, Pasadena, CA 91 125, USA}
\altaffiltext{6}{Universitat Wien, Institut fur Astrophysik, T\"urkenschanzstrasse 17, 1180 Wien, Austria}
\altaffiltext{7}{Department of Physics and Astronomy, University of Canterbury, Private Bag 4800, Christchurch, 8140, New Zealand}
\altaffiltext{8}{Institute for Astronomy, Royal Observatory Edinburgh, Blackford Hill, Edinburgh, EH9 3HJ, UK.}
\altaffiltext{9}{School of Physics and Astronomy, Cardiff University, Queen's Buildings, The Parade, Cardiff, CF24 3AA, UK}
\altaffiltext{10}{Institut d'Astrophysique de Paris, UMR7095 CNRS-Universite Pierre et Marie Curie, 98bis bd Arago, F-75014 Paris, France}
\altaffiltext{11}{National Radio Astronomy Observatory, 520 Edgemont Rd, Charlottesville, VA, 22903, USA}
\altaffiltext{12}{European Southern Observatory, Karl-Schwarzschild-Str. 2, Garching, Germany}
\altaffiltext{13}{INAF, Osservatorio Astronomico di Padova, Vicolo Osservatorio 5, I-35122 Padova, Italy}
\altaffiltext{14}{Leiden Observatory, Leiden University, P.O. Box 9513, NL-2300 RA Leiden, The Netherlands}
\altaffiltext{15}{Joint ALMA Observatory, Alonso de Cordova 3107, Vitacura, Santiago, Chile}
\altaffiltext{16}{European Southern Observatory, Alonso de Cordova 3107, Vitacura, Santiago, Chile}

\begin{abstract}
We exploit long-baseline ALMA sub-mm observations of the lensed
star-forming galaxy SDP\,81 at $z$\,=\,3.042 to investigate the
properties of inter-stellar medium on scales of 50--100\,pc.  The
kinematics of the $^{12}$CO gas within this system are well
described by a rotationally-supported disk with an
inclination-corrected rotation speed, $v_{\rm
  rot}$\,=\,320\,$\pm$\,20\,km\,s$^{-1}$ and a dynamical mass of
M$_{\rm
  dyn}$\,=\,(3.5\,$\pm$\,1.0)\,$\times$\,10$^{10}$\,M$_{\odot}$
within a radius of 1.5\,kpc.  The disk is gas rich and unstable,
with a Toomre parameter, $Q$\,=\,0.30\,$\pm$\,0.10 and so collapse
in to star-forming regions with Jeans length $L_{\rm
  J}\sim$\,130\,pc.  We identify five star-forming regions within
the ISM on these scales and show that their scaling relations
between luminosity, line-widths and sizes are significantly offset
from those typical of molecular clouds in local Galaxies (Larson's
relations).  These offsets are likely to be caused by the high
external hydrostatic pressure for the interstellar medium (ISM),
$P_{\rm tot}$\,/\,$k_{\rm
  B}\sim$\,$40_{-20}^{+30}$\,$\times$\,10$^{7}$\,K\,cm$^{-3}$, which
is $\sim$\,10$^4$\,$\times$ higher than the typical ISM pressure in
the Milky Way.  The physical conditions of the star-forming ISM and
giant molecular clouds appears to be similar to the those found in
the densest environments in the local Universe, such as those in the
Galactic center.
\end{abstract}

\keywords{galaxies: starburst; galaxies: evolution; galaxies: high-redshift}

\section{Introduction}


Giant molecular clouds in local galaxies follow well known scaling
relations between CO velocity line width ($\sigma$) and their physical
extent, $R$, with $\sigma\propto R^{1/2}$ and mean molecular gas
density scales and size; $\langle n({\rm H}_2)\rangle \propto R^{-1}$
\citep{Larson81,Bolatto06}.  These scalings reflect the dynamical
state of the turbulent molecular gas in the inter-stellar medium
(ISM).  Since most of the stars in local, massive spheroids and
elliptical galaxies appear to have formed early in the history of the
Universe ($z\sim$\,2--3), examining the physical, dynamical and
thermal state of the molecular gas within the ISM of galaxies at this
epoch acquires special importance.  However, to measure the properties
of individual star-forming regions requires a spatial resolution of at
least $\sim$\,100\,pc (sufficient to resolve sizes and velocity
dispersions of the most massive giant molecular clouds).  To date,
this has only been achieved in a few rare examples of high-redshift
galaxies whose images have been gravitationally lensed by massive
galaxy clusters \citep[e.g.\ ][]{Jones10,Livermore12a,Livermore15}.

To test whether the scaling relations that govern the structure of
local GMCs are valid in the dense and rapidly evolving ISM of
high-redshift, gas-rich galaxies, in this {\it Letter} we exploit ALMA
observations of SDP\,81 -- a star-forming galaxy at $z$\,=\,3.042
whose image has been gravitationally lensed by a factor
15.8\,$\pm$\,0.7$\times$ by a massive, foreground ($z\sim$\,0.299)
galaxy \citep{Dye15}.  The combination of long-baselines, together
with gravitational lensing means we are able to resolve the largest
giant molecular clouds with in the ISM \citep{Solomon87,Scoville89} on
scales approaching 50\,pc.  We use the spatially resolved 1.0\,mm
(rest-frame 250\,$\mu$m) continuum imaging to identify the brightest
star-forming regions, and measure their sizes, luminosities and
velocity dispersions.  We use a $\Lambda$CDM cosmology with
H$_0$\,=\,72\,km\,s$^{-1}$\,Mpc$^{-1}$, $\Omega_m$\,=\,0.27 and
$\Omega_{\Lambda}$\,=\,1$-\Omega_m$ \citep{Spergel04} and a Chabrier
IMF.

\begin{figure*}
\centerline{
  \psfig{figure=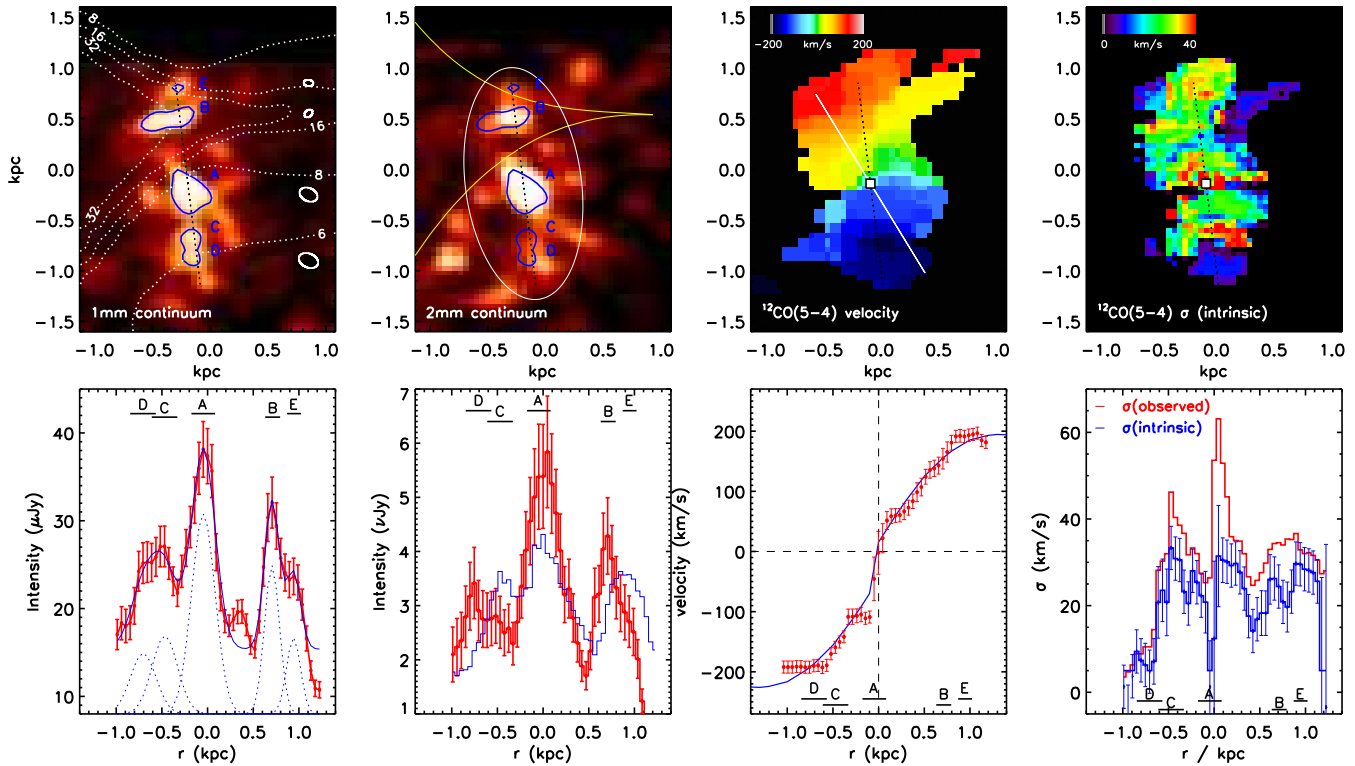,width=4in,angle=90}}
\caption{Source plane images of SDP\,81. {\it Top Row: Left:}
  Source-plane 1.0\,mm (rest-frame 250$\mu$m image) continuum image of
  SDP\,81.  The star-forming clumps, A--E, are identified by the solid
  contours.  The black dashed line defines the major morphological
  axis and the white contours indicate the amplification map.  The
  luminosity weighted amplification at 1.0\,mm is 16.0\,$\pm$\,0.7.
  The white ellispes show the source-plane PSF at the location of the
  clumps (offset to the right hand size of the image for clarity).
  {\it Center Left:} Source-plane 2.1\,mm image with the clumps
  identified at 1.0\,mm also highlighted.  The solid line denotes the
  caustic curve from the best-fit lens model.  {\it Center Right:}
  $^{12}$CO(5--4) velocity field.  The gas disk has an observed
  peak-to-peak velocity gradient of 210\,$\pm$\,10\,km\,s$^{-1}$.  The
  solid white line shows the major kinematic axis and the dotted line
  denotes the major 1.0\,mm morphological axis.  {\it Top Right:}
  Beam-corrected $^{12}$CO(5--4) line-of-sight velocity
  dispersion. {\it Bottom Row: Left:} One-dimensional profile of the
  1.0\,mm continuum emission extracted across the major morphological
  axis.  We label the five bright star-forming regions (A--E).  The
  best-fit profiles are shown as dashed lines.  {\it Center Left:}
  One-dimensional profile of the 2.1\,mm intensity extracted along the
  major axis of the 1.0\,mm continuum with the $^{12}$CO(5--4)
  intensity profile overlaid for comparison (blue). {\it Center
    Right:} One-dimensional velocity profile of the gas disk with
  best-fit dynamical model overlaid.  {\it Right:} One-dimensional
  velocity dispersion profile (red) and intrinsic, beam-corrected
  ($\Delta$V\,/\,$\Delta$r) velocity dispersion profile (blue).  In
  all of the lower panels we show the positions of the star-forming
  regions.  The source-plane sub-mm morphology appears
  complex\,/\,clumpy, but with the star-forming regions embedded in a
  dense, rotating disk.}
\label{fig:recon}
\end{figure*}

\section{Observations and Source Plane Reconstruction}

SDP\,81 was identified from the {\it H-ATLAS} survey as a bright
sub-mm source at $z$\,=\,3.042 by \citet{Negrello10}.  Optical imaging
and spectroscopy also revealed the presence of a massive foreground
galaxy at $z$\,=\,0.299 which lenses the background galaxy
\citep{Negrello14}.  Observations of SDP\,81 with ALMA in its long
baseline configuration (up to 15\,km) were taken in 2014 October.
These observations and reduction are described by \citet{Vlahakis15}.
Briefly, the ALMA Band\,7 (1.0\,mm) continuum observations have a
resolution of 31\,$\times$\,23\,mas and reach a 1-$\sigma$ depth of
11\,$\mu$Jy\,/\,beam.  At the redshift of the galaxy, these
observations sample the rest-frame 250$\mu$m emission.
Lower-frequency observations of the $^{12}$CO(5-4) and continuum
emission at 2.1\,mm were also made, reaching a resolution of
56\,$\times$\,50\,mas.

\citet{Dye15} construct a detailed lens model for the system using
both the ALMA sub-mm and \emph{Hubble Space Telescope; (HST)} imaging  \citep[see also][]{Rybak15}.
The best-fit lens model suggests that the background source comprises
three dominant components, a dense gas disk (that lies inside the
caustic and so gives rise to the bright sub-mm emission in the image
plane) and two galaxy nuclei (which lie on - or just outside - the
caustic) which are visible in the \emph{HST} $JH$-band
imaging.  Dye et al.\ interpret the complex morphology as a merging
system in which the gas disk is a result of an early stage
interaction.  
In their model, the luminosity weighted amplification is $\mu_{\rm
  submm}$\,=\,15.8\,$\pm$\,0.7 and $\mu_{\rm
  opt}$\,=\,10.2\,$\pm$\,0.5 for the sub-mm and rest-frame optical
emission respectively.  Thus, correcting for lensing amplification,
the observed 850\,$\mu$m flux density of the galaxy is S$_{\rm 850\mu
  m}\sim$\,1.2\,mJy, which is representative of the SMG population
which has recently been studied in detail, in particular with ALMA
\citep{Hodge13,Karim13,Simpson14}.  However, the amplifications means
that the average source-plane resolution is $\sim$\,50--100\,pc -- a
factor $\sim$\,30\,$\times$ higher than that so-far achieved in the
non-lensed case \citep[e.g\ ][]{Simpson15a,Ikarashi15}.

\section{Analysis \& Discussion}

Using the lens modeling from \citet{Dye15}, we reconstruct the
source-plane morphology of SDP\,81 at 1.0\,mm and 2.1\,mm and show
these in Fig.~\ref{fig:recon}.  The dust continuum morphology appears
clumpy, and we isolate five of the brightest star-forming regions from
the highest-resolution (1.0\,mm) image (using the criteria that they
are 5-$\sigma$ above the {\it local} background), and label these
A--E.  Independently, we also use the {\sc clumpfind} algorithm
\citep{Williams94}, which isolates clumps A--D, although misses clump
E.  However, none of our conclusions significantly change if we
include\,/\,exclude region E from the analysis below.

To measure the source-plane point spread function (PSF) we reconstruct
the beam in a grid of positions in the image plane and measure the PSF
at each reconstructed position in the source-plane.  On average, the
source-plane PSF has a FWHM\,$\sim$\,60\,pc (Fig.~\ref{fig:recon}).
In Fig.~\ref{fig:recon} we also show the one-dimensional 1.0\,mm,
2.1\,mm and $^{12}$CO(5--4) emission profiles extracted from the major
morphological axis of the source.  The dust continuum and $^{12}$CO
emission profiles are not perfectly aligned.  However, at this
resolution, offsets between the CO-emitting gas and continuum may be
expected in regions with high star-formation density where the
$^{12}$CO(5-4) (which traces the warm and dense gas) is shock-heated,
which efficiently raises the gas temperature and density through
mechanical heating but does not heat the dust.  Indeed, if the disk
fragments into a number of large star forming regions (as we discuss
below), we expect large-scale shocks where the complexes interact.

In Fig.~\ref{fig:recon} we also show the source-plane $^{12}$CO(5--4)
velocity field \citep[see also][]{Dye15} which resembles a rotating
system with a peak-to-peak velocity of 210\,$\pm\,$10\,km\,s$^{-1}$
within 1.5\,kpc.  The best-fit disk model suggests
an inclination of $\theta$\,=\,40\,$\pm$\,5$^\circ$ and so a dynamical
mass of M$_{\rm
  dyn}$\,=\,(3.5\,$\pm$\,1.0)\,$\times$\,10$^{10}$\,M$_{\odot}$ within
a radius of 1.5\,kpc.  The total gas mass for the disk, estimated from
either the (amplification corrected) $^{12}$CO(1--0) luminosity, or
using the far-infrared SED and an appropriate dust-to-gas ratio is
2.7--3.9\,$\times$\,10$^{10}$\,M$_{\odot}$ \citep{Dye15}.  Together
these indicate that the central regions of the disk are baryon
dominated with a gas fraction of $f_{\rm gas}\sim$\,M$_{\rm
  gas}$\,/\,M$_{\rm dyn}\sim$\,70--90\%.


\begin{figure*}
\centerline{
  \psfig{figure=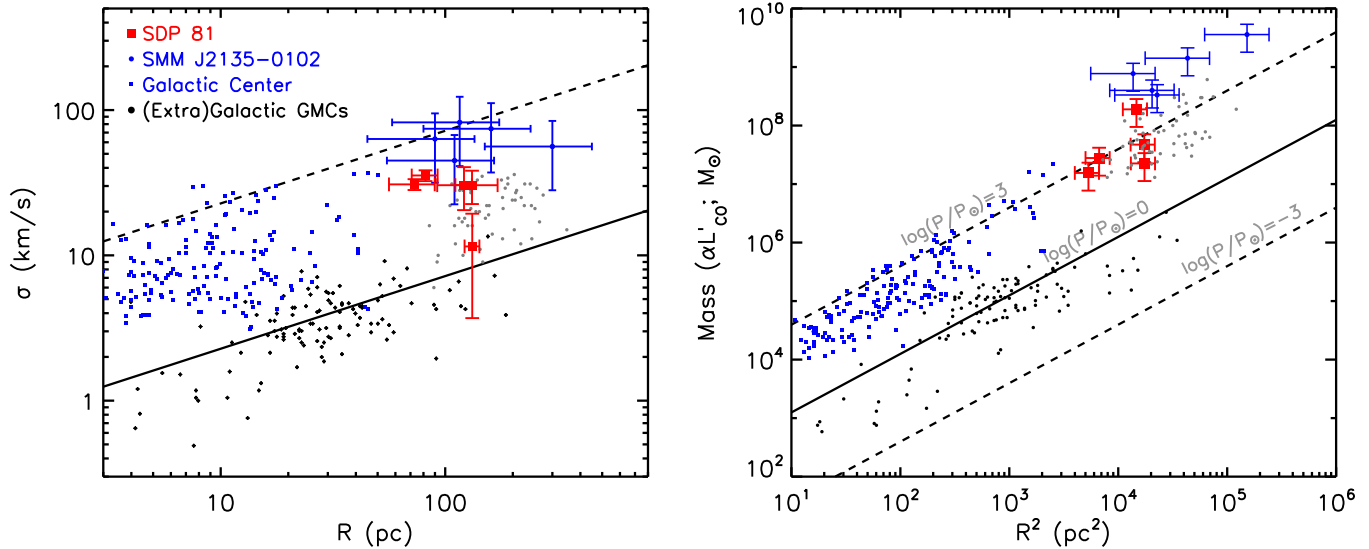,width=7.2in,angle=0}}
\caption{The relations between size, luminosity and velocity
  dispersions for the star-forming regions in SDP\,81 compared to
  those in the both quiescent- and more extreme- evironments of the
  Milky Way and other local galaxies {\it Left:} The velocity
  dispersion versus cloud radius. The solid line shows the
  line-width--size relation for local GMCs \citep{Larson81,Bolatto06}.
  The normalisation of this scaling relation is set by the gas
  pressure in the disk \citep{Elmegreen89}.  The dashed line shows the
  local relation but offset by a factor of 10\,$\times$.  GMCs in high
  pressure environments, such as the Galactic Center (blue points) or
  in gas rich galaxies such as M\,64 (grey points) are offsets in the
  sense that they have higher velocity dispersions at fixed size.
  Indeed, the star-forming regions in SDP\,81 appear to have velocity
  dispersions that are $\sim 4\times$ higher than expected.  However,
  it is also interesting to note that their velocity dispersions are
  $\sim$\,2\,$\times$ lower than the star-forming regions in
  SMM\,J2135$-$0102.  {\it Right}: The relation between gas mass and
  size; $M({\rm H}_2)-R^2$.  The star-forming regions in SDP\,81 are
  again offset from those of local GMCs, and instead are consistent
  with having much higer luminosities at fixed size -- a consequence
  of the high turbulent pressure of the ISM (see \S~3).  }
\label{fig:LRS}
\end{figure*}

The Toomre parameter, $Q$ characterises the stability of a disk
against local axisymmetric perturbations.  Gas rich disks with $Q<1$
should fragmement and collapse in to star forming regions.  Numerical
simulations have also suggested that tidal interactions and dynamical
friction should force the resulting star-forming regions towards the
center of the galaxy, where they should coallesce to form a bulge
(which in turn stabilises the disk against further collapse;
e.g.\ \citealt{Ceverino10,Genzel08}).  The Toomre parameter is
calculated by $Q$\,=\,$\sigma_{\rm r}\kappa$\,/\,$\pi G \Sigma_{\rm
  gas}$ where $\kappa$\,=\,$a$\,V$_{\rm max}$\,/\,$R$ is the epicycic
frequency (with $a$\,=\,$\sqrt3$), $\sigma_{\rm r}$ is the line of
sight velocity dispersion; and $\Sigma_{\rm gas}$ is the mass surface
density of the gas \citep{Toomre64}.  For SDP\,81, we derive
$Q$\,=\,0.30\,$\pm$\,0.10, which is lower than typical for the gas in
local ULIRGs ($Q\sim$\,1; \citealt{Downes98}), and slightly lower than
the average Toomre $Q$ of other gas-rich star-forming $z\sim$\,2
($<Q>$\,=\,0.85\,$\pm$\,0.13; \citealt{Genzel14} -- although their
sample also extend down to comparable values,
$Q$\,=\,0.18\,$\pm$\,0.02).


This global instability will cause large, dense gas clouds to form in
the molecular gas.  The fragmentation of the gas occurs on scales of
the Jeans length, L$_{\rm J}$ which can be estimated using the gas
surface density and average gas velocity dispersion according to
L$_{\rm J}$\,=\,$\pi\sigma_{\rm r}^2/8 G \Sigma_{\rm gas}$.  For
SDP\,81, with $\Sigma_{\rm
  gas}$\,=\,4\,$\pm$\,1\,$\times$\,10$^9$\,M$_{\odot}$\,kpc$^2$ and
$\sigma_{\rm r}$\,=\,30--35\,km\,s$^{-1}$ we estimate L$_{\rm
  J}$\,=\,130$_{-60}^{+200}$\,pc.  If the gas fragments on these
scales, the corresponding mass of the clouds that form should be
$\sigma_{\rm r}^4/G^2\Sigma_{\rm
  gas}\sim$\,1--2\,$\times$\,10$^8$\,M$_{\odot}$.

Given our source-plane resolution is at least comparable to the Jeans
length, we investigate the properties of the star-forming regions
on these scales.  As Fig.~\ref{fig:recon} shows, the disk
contains several bright star-forming regions (A-E)
and we measure their size, luminosity and velocity dispersion (using
the beam-corrected velocity dispersion map for the velocity
dispersions; \citealt{Davies11}).  In all five regions the
$^{12}$CO(5--4) has an intrinsic velocity dispersions of
10--35\,km\,s$^{-1}$ (Table~1).  To estimate sizes for the molecular
gas associated with these regions, we use two approaches.  First, we
fit the sub-mm light profile with Gaussian profiles (centered on each
star-forming region; Fig.~\ref{fig:recon}).  Second, we measure the
area subtended by a contour which is 5-$\sigma$ above the local
background.  The resulting sizes (deconvolved for the source-plane
PSF) are FWHM\,=\,170--310\,pc, which is comparable to the {\it
  initial} fragmentation scale.
We also estimate the molecular gas mass in the vicinity of the
star-forming regions using the $^{12}$CO(5--4) line luminosity and
accounting for both the local background and the
$^{12}$CO(1-0)\,/\,$^{12}$CO(5-4) luminosity ratio in velocity bins
\citep[see][]{Dye15}, obtaining gas masses of M$_{\rm
  gas,clumps}$\,=\,1--5\,$\times$\,10$^{8}$\,M$_{\odot}$.  These
masses are consistent with those derived from those using their line
width and sizes with M$_{\rm dyn}$\,=\,$C\sigma^2$\,R/$G$ with
$C$\,=\,5 -- as approriate for a uniform density sphere -- with
M$_{\rm dyn}$\,/\,M$_{\rm CO}$\,=\,1.5\,$\pm$\,0.5).

In Fig.~\ref{fig:LRS} we compare the velocity dispersion, luminosity
and sizes of the star-forming regions in SDP\,81 with similar
measurements for GMCs in the local Universe (compiled from
observations of the Milky-Way disk and other quiescent galaxies;
\citealt{Bolatto06}), as well as to the properties of GMCs and
star-forming regions in more extreme environments, including the
galactic center and the ISM of SMM\,J2135$-$0102 -- a star-forming
galaxy at $z$\,=\,2.32 where similar measurements have been made
\citep{Swinbank11}.  At a fixed size, the velocity dispersions of the
star-forming regions in SDP\,81 appear to be $\sim$\,4$\times$ larger
than those of GMCs in quiescent environments, but more similar to
those found in the Galactic Center and other gas-rich environments.
In the local Universe, these offsets are usually attributed to high
external pressures on the cloud surfaces due to high gas densities
\citep{Keto05, Blitz04, Blitz06}.

To interpret the offsets in the properties of the star-forming regions
in SDP\,81, we use the dynamics and surface density of the molecular
gas to estimate the ISM pressure (the kinetic pressure resulting from
non-ordered mass motions).  In a rotating gas disk, the
mid-plane hydrostatic pressure is given by:

\begin{equation}
  \rm P_{\rm tot} \approx \frac{\pi}{2}G \Sigma_{\rm gas}\left[\Sigma_{\rm gas}+\left(\frac{\sigma_{\rm gas}}{\sigma_{\star}}\right)\Sigma_{\star}\right]
  \label{eqn:Ptot}
\end{equation}

\noindent where $\Sigma_{\rm gas}$, $\Sigma_{\star}$ and $\sigma_{\rm
  gas}$, $\sigma_{\star}$ are the surface density and velocity
dispersion of the gas and stars respectively.  In the Milky Way,
$P_{\rm tot}/k_{\rm B}\sim 1.4\times 10^4$\,cm$^{-3}$\,K
\citep{Elmegreen89}.  The stellar mass of SDP\,81 estimated from the
rest-frame UV\,/\,optical photometry is
M$_{\star}$\,=\,(6.6$^{+2.6}_{-1.9}$)\,$\times$\,10$^{10}$\,M$_{\odot}$
\citep{Negrello14}, although the stars appear to be offset by
$\sim$\,1.5\,kpc from the dust and gas emission \citep{Dye15}, and so
we adopt M$_{\star}\lsim$\,3\,$\times$\,10$^{10}$\,M$_{\odot}$ as an
upper limit on the stellar contribution within the gas disk.  We also
assume that the velocity dispersion of the gas and stars are
comparable, $\sigma_{\rm gas}$\,/\,$\sigma_{\star}\sim$\,1 (although
allow this to vary from 0.5--2 in the calculation below).  Given the
high gas surface density, the mid-plane hydrostatic pressure is high,
$P_{\rm tot}$\,/\,$k_{\rm
  B}\sim$\,$40_{-20}^{+30}$\,$\times$\,10$^{7}$\,K\,cm$^{-3}$ and
although we caution this value has considerable uncertainty, this
pressure is $\gsim$\,10$^4$\,$\times$ higher than the typical pressure
in the Milky Way disk ($10^{4}$\,K\,cm$^{-3}$).  This pressure is also
$\sim$\,10\,$\times$ higher than inferred for the ISM in more extreme
environments, such as in the Galactic Center or the Atennae
\citep{Rosolowski05,Keto86,Wilson03}.  However, the compact disks of
some ULIRGs, with high gas surface densities ($\ga 5\times
10^3$\,M$_{\odot}$\,pc$^{-2}$, \citealt{Downes98}) may also result
comparably high pressures.  Finally, we note that the implied pressure
in SDP\,81 is compatible with recent hydro-dynamic models which
suggest that the typical pressure in the ISM of star-forming galaxies
in the should increases from $\sim$\,10$^4$\,K\,cm$^{-2}$ at
$z$\,=\,0.1 to $\sim$\,10$^6$--10$^{7}$\,K\,cm$^{-2}$ at $z$\,=\,2,
reaching $\sim$\,10$^9$\,K\,cm$^{-2}$ in some systems \citep{Crain15}.

%
%
\begin{table*}
\begin{center}
\caption{Clump Properties}
\begin{tabular}{lccccccc}
\hline
\hline
ID           &  $v_{\rm clump}$    &  Amplification      & FWHM$_{\rm clump}$   & $\sigma_{\rm clump}$ &  $f_{\rm CO(5-4)}$   & $r_{\rm 54}$    & M$_{\rm gas}$                   \\
             &  (km\,$s^{-1}$)    &  ($\mu$)            & (pc)            & (km\,$s^{-1}$)      &  (mJy\,km\,s$^{-1}$)  &               & ($\times$\,10$^{8}$\,M$_{\odot}$) \\
\hline
A            &  $-$42\,$\pm$\,65  &  7.2\,$\pm$\,0.8   &  282\,$\pm$\,25  &  30\,$\pm$\,9      &   57\,$\pm$\,4  & 0.28\,$\pm$\,0.05   & 1.2 \\
B            &    162\,$\pm$\,12  & 40.1\,$\pm$\,1.0   &  188\,$\pm$\,25  &  35\,$\pm$\,3      &   24\,$\pm$\,4  & 0.30\,$\pm$\,0.12   & 2.8 \\
C            & $-$150\,$\pm$\,28  &  6.2\,$\pm$\,0.18  &  305\,$\pm$\,95  &  31\,$\pm$\,7      &   46\,$\pm$\,4  & 0.28\,$\pm$\,0.08   & 4.8 \\
D            & $-$176\,$\pm$\,2   &  5.9\,$\pm$\,0.15  &  300\,$\pm$\,70  &  11\,$\pm$\,7      &   35\,$\pm$\,4  & 0.36\,$\pm$\,0.05   & 0.8 \\
E            &    204\,$\pm$\,5   & 30.4\,$\pm$\,7.4   &  170\,$\pm$\,40  &  31\,$\pm$\,3      &   28\,$\pm$\,4  & 0.18\,$\pm$\,0.04   & 0.7 \\
\hline
\label{table:clump_props}
\end{tabular}
\end{center}
\noindent{\footnotesize Notes: $v_{\rm clump}$ denotes the velocity of
  the disk at the position of the star-forming region with respect to
  the systemic redshift.  The amplifications are 1.0\,mm emission
  luminosity weighted values.  FWHM$_{\rm clump}$ have been
  deconvolved for the source-plane PSF.  $\sigma_{\rm clump}$ is the
  velocity dispersion of the clump as measured from the intrinsic
  $^{12}$CO(5--4) velocity dispersion map.  $r_{\rm 54}$ is the ratio
  of the $^{12}$CO(5--4)\,/\,$^{12}$CO(1--0) luminosities derived from
  velocity and amplification maps; \citep{Dye15}.  M$_{\rm gas}$
  denotes the gas mass of each clump assuming M$_{\rm
    gas}$\,=\,$\alpha_{\rm CO}$\,L$'_{\rm CO(1-0)}$ with L$'_{\rm
    CO(1-0)}$\,=\,L$'_{\rm CO(1-0)}$\,/\,r$_{\rm 54}$ and $\alpha_{\rm
    CO(1-0)}$\,=\,M$_{gas}$\,/\,L$'_{\rm CO(1-0)}$\,=\,1.

}
\end{table*}

To relate the pressure to the properties of the star-forming regions,
we use ``Larson's relations'' for turbulent molecular clouds.
Following \citet{Elmegreen89}, the velocity-dispersion--size and
mass-size relations can be cast as:

\begin{equation}
  \sigma = \sigma_{\circ} \left(\frac{P_{\rm ext}/k_{\rm B}}{10^4\,{\rm
      K\,cm}^{-3}}\right)^{1/4}\left(\frac{R}{\rm pc}\right)^{1/2}
  \label{eqn:Larson2}
\end{equation}
and
\begin{equation}
  M({\rm H}_2)=290\left(\frac{P_{\rm ext}/k_{\rm B}}{10^4\,{\rm cm}^{-3}\,{\rm K}}\right)^{1/2}\left(\frac{R}{\rm pc}\right)^2\,{\rm M}_{\odot}.
  \label{eqn:MH2}
\end{equation}
with $\sigma_\circ$\,=\,1.2\,km\,s$^{-1}$ \citep{Larson81}.  Using
Equation~\ref{eqn:Larson2}, a pressure of 10$^4$\,$\times$ that of the
Milky-Way therefore suggests that the velocity dispersions of the
clouds within the ISM of SDP\,81 should be $\gsim$\,10$\times$ those
in the Milky Way respectively (at a fixed size).  The average velocity
dispersion of the clumps in SDP\,81 is four times larger at a fixed
size than predicted from GMCs in the Milky-Way, and although this is
lower than the factor of 10$\times$ predicted for pressure induced
offsets alone, we reiterate that this is a simplified model that
nevertheless relates the offsets in the scaling relations for GMCs
with the gas densities and pressures in the ISM.  Finally, in
Fig.~\ref{fig:LRS} we plot the mass--radius relation for a similar
range of quiescent- and extreme- environments in both the Milky Way
and other nearby galaxies and plot the positions of star-forming
regions in SDP\,81, which are again offset to higher masees by a
factor $\sim$\,10\,$\times$ (at a fixed size) with respect to those of
GMCs in quiescent environments.  This can also be attributed to the
high turbulent pressure (Equation~\ref{eqn:MH2}).  However, it is
interesting to note that they do not appear as massive as the
star-forming regions in SMM\,J2135$-$0102, although the latter is a
natural consequence of the mass scale for collapse within the disk
given their relative velocity dispersions and gas surface densities
($\sigma_{\rm SMM\,J2135}$\,/\,$\sigma_{\rm SDP81}\sim$\,2.3 and
$\Sigma_{\rm gas,SMM\,J2135}$\,/\,$\Sigma_{\rm gas,SDP81}\sim$\,2)
which results in a mass scale difference for the clumps of M$_{\rm
  cl,SMMJ2135}$\,/\,M$_{\rm cl,SDP81}\sim$\,25 -- consistent with
Fig.~\ref{fig:recon}.

\section{Conclusions}

Using long-baseline ALMA observations, we have mapped the distribution
of star-formation and molecular gas in the lensed, star-forming
$z$\,=\,3.042 galaxy, SDP\,81, on physical scales of
$\sim$\,50--100\,pc.  The $^{12}$CO(5--4) dynamics suggest that the
molecular gas is located in a disk with an inclination-corrected
rotation speed of $v_{\rm rot}$\,=\,320\,$\pm$\,20\,km\,s$^{-1}$ and a
dynamical mass of M$_{\rm
  dyn}$\,=\,(3.5\,$\pm$\,1.0)\,$\times$\,10$^{10}$\,M$_{\odot}$ within
a radius of 1.5\,kpc.

The gas disk appears to be Toomre unstable, $Q$\,=\,0.30\,$\pm$\,0.10.
This instability will cause large, dense star-forming regions to
collapse on scales of the Jeans length, $L_{\rm
  J}$\,=\,130$_{-60}^{+200}$\,pc.  We identify five star-forming
regions on these scales in the rest-frame 250-$\mu$m continuum and
measure their size, luminosity and CO velocity dispersions.  We show
that these star-forming regions do not lie on the local relations for
GMCs, but are instead systematically offset such that the velocity
dispersion is $\sim$\,4\,$\times$ higher than typical GMCs at a fixed
size.

The gas dynamics and surface density suggest that the ISM should be
highly pressurised, and we estimate a mid-plane hydrostatic pressure
which is $\sim$\,10$^4$\,$\times$ higher than typically found in the
Milky Way.  These high pressures appear to be responsible for the in
offsets in the scaling relations of the star-forming regions compared
to those of typical GMCs in galaxies in the local Universe.  Within
the star-forming ISM of this dense gas disk, the physical conditions
appear to be similar to those only seen in the densest environments in
the local Universe \citep[e.g.\ ][]{Kruijssen13}

\section*{Acknowledgments}

We would like to thank the anonymous referee for a constructive report
on this paper.  AMS acknowledges an STFC Advanced Fellowship
(ST/H005234/1) and the Leverhume foundation.  IRS acknowledges STFC
(ST/I001573/1), the ERC Advanced Grant DUSTYGAL 321334 and a Royal
Society/Wolfson Merit Award.  CF acknowledges CAPES funding
(proc. 12203-1).  LD, RJI, and IO acknowledge support from the
European Research Council (ERC) in the form of Advanced Grant, {\sc
  cosmicism}.  We also thank Padelis Papadopoulos for useful
discussions.  This paper uses data from ALMA program
ADS/JAO.ALMA\#2011.0.00016.SV.  ALMA is a partnership of ESO, NSF
(USA), NINS (Japan), NRC (Canada), NSC and ASIAA (Taiwan), and KASI
(Republic of Korea) and the Republic of Chile. The JAO is operated by
ESO, AUI/NRAO and NAOJ.


\end{document}